\documentclass[12pt]{iopart}

\usepackage{graphicx,xcolor}

\begin{document}

\title[]{The rise and fall of hubs in Self-Organized Critical learning networks} 

\author{Anjan Roy $^{1,\dagger}$, Serena Di Santo\,$^{2,\dagger}$,  and Matteo Marsili $^{3}$}

\address{
$^{1}$ Department of Industrial Engineering and Management, Ben-Gurion University of the Negev,
Beer Sheva, Israel\\
$^{2}$ Center for Theoretical Neuroscience, Columbia University, NY, USA \\
$^{3}$ Quantitative Life Science Section, The Abdus Salam International Centre for Theoretical Physics, Trieste, Italy \\
$^{\dagger}$ Equal contribution \\}
\ead{anjan@post.bgu.ac.il}
\vspace{10pt}

\begin{abstract}
Information processing networks are the result of local rewiring rules. In many instances, such rules promote links where the activity at the two end nodes is positively correlated. The conceptual problem we address is what network architecture prevails under such rules and how does the resulting network, in turn, constrain the dynamics. We focus on a simple toy model that captures the interplay between link self-reinforcement and a Self-Organised Critical dynamics in a simple way. Our main finding is that, under these conditions, a core of densely connected nodes forms spontaneously. Moreover, we show that the appearance of such clustered state can be dynamically regulated by a fatigue mechanism, eventually giving rise to non-trivial avalanche exponents.
\end{abstract}

%
\vspace{2pc}
\maketitle
%
%

Since Bak's proposal \cite{BakBrain1,BakBrain2}, experimental evidences that the Self-Organised Criticality (SOC) paradigm applies to neural dynamics have been accumulating \cite{BeggsPlenz}. Indeed, as observed in \cite{HerzHopfield} the threshold dynamics that lies at the origin of SOC behaviour, is similar to that of integrate-and-fire neurons. The same threshold dynamics 
has proven insightful in characterizing information processing and spreading behaviour in a much wider plethora of phenomena, including belief formation, information and opinion exchange, as well as adoption cascades, in social networks  \cite{Watts}. 

Besides the generic SOC phenomenology, the performance of these networks as information processing or aggregation devices, depends in crucial ways on the specific wiring of the network. For example, Ref. \cite{Basset} highlights how information transmission in models of neural activity depends crucially on the presence of network motifs and Ref. \cite{Livan} shows how, the presence of cliques in a social network is responsible for the failure of information aggregation. Yet, in both cases, the prevailing network is the result of local dynamical rules: the connections between two neurons can strengthen, weaken or even decay based on their relative spike-timing, a well-studied phenomenon referred to as Hebbian plasticity \cite{Hebb}. The more two neurons spike together, the more the link between them will be reinforced. 

Likewise, the fate of social links depends in many ways on the quantity and the quality of exchanges that occur between the agents. Changes in the network's structure typically occur on time scales that are much longer than the one of the dynamics of processes that take place on the network. Hence, the long term properties of these systems ultimately depend on the working of local connectivity adjustment processes. The key question, then, is what local network dynamics can promote specific properties in the long term?

Related work \cite{shinkim2006} reports that spike timing dependent synaptic plasticity (STDP) -- a biologically realistic mechanism that implements the Hebbian principle-- reorganizes a globally connected network of excitable units into a functional network with a broad degree distributions. Moreover, other related studies which consider coupled map networks whose links are rewired based on correlation in the dynamics of the maps \cite{jost2009} and using more realistic spiking neuronal models with biologically motivated rewiring rules \cite{ren2010} have shown the emergence of specific network motifs similar to those seen in brain networks.

This paper tackles the problem of understanding the interplay between the dynamics on and of a network from an abstract perspective, building upon the well known self-organizing dynamics of the sandpile model and adding a biologically inspired rewiring rule, but abstaining from detailing the distinctiveness of any biological or social system. Indeed, we address a simpler and more specific question, namely, how a local reinforcement rule on the links, on top of a SOC dynamics in the nodes, shapes the network globally. More precisely, a link gets reinforced when the activity of the up-stream node triggers that of the down-stream one, and it weakens if the downstream node stays idle. The choice of an over-simplified model is intentional and aims at identifying robust statistical behaviours that can be fully characterised.

The model assigns to each directed link a variable that quantifies the co-activation of the nodes at its ends. Links with poor co-activation are rewired to random nodes. When all links where this variable is smaller than a threshold are rewired, the model undergoes a phase transition as a function of the threshold. For low values of the threshold, we find a ``solid'' phase, where most links are permanent and rewiring is absent or rare. For large values of the threshold, on the contrary, most links are constantly rewired and the network is in a ``liquid state''. In both cases, the resulting network is an homogeneous random graph. In between those two regimes we observe an intermediate phase characterized by the emergence of a clique of densely connected nodes called hubs. Extremal link-rewiring dynamics, where the weakest link only is rewired, allows us to zoom into the critical state and fully characterise it. As in other SOC models \cite{BakSneppen}, extremal dynamics embodies a sharp separation of time-scales between the (fast) dynamics on the network and the (slow) evolution of the network itself. Here we find that the network ``condenses'' into a state characterised by a clique of hubs, where most of the activity concentrates, in a sea of peripheral nodes that merely provides input to the clique. This conclusion is found to be robust with respect to a variety of modifications of the model. Finally, if the nodes are endowed with a fatigue-like mechanism, these hubs can disappear and eventually also re-appear, making the network more plastic and giving rise to non-trivial dynamical behaviours. This sort of regulatory mechanisms are very common in a variety of systems and in particular the one that we introduce here is inspired by homeostatic processes (e.g. adaptation, synaptic depression {\cite{tsodyks1997,abbott1997,lee2019mechanisms})} that maintain the dynamic range of neural networks around a point of moderate activity.

The rest of the paper is organised as follows:
in Section 1 we formalize the activity propagation and rewiring rules mentioned above, as we introduce the Parametric Model (where the rewiring rule is based on the value of a threshold), 
the Extremal Model (where only the weakest link is rewired and the system self-organizes to the critical state) 
 and the {Adaptive} Model (where fatigue is added). In Section 2 we show the macroscopic properties resulting from the simulation of the three models. First, in the Parametric Model, we characterize the phase transition generated by the reinforcement rule. Then we show how the extremal link-rewiring dynamics allows the network to spontaneously reach a steady state corresponding to the critical point, with no need for parameter tuning. Next, we illustrate how the introduction of fatigue disrupts the hubs, giving rise to a more heterogeneous network structure and dynamical behaviour. Finally, we discuss the results, with a focus on the biological network features that this drastically simplified framework can contribute to unveil.
 
\section{The Models}

\subsection{Dynamics on the Network: Sandpile Model on a Random Network}
\label{sec1}

Let $G$ be a directed graph with $N$ sites and adjacency matrix $g_{i,j}=1$  if link $i\to j$ exists and $g_{i,j}=0$ otherwise, for all $i\neq j$ ($g_{i,i}=0$). Let there be $K_{out}^i$ outgoing links for site $i$, that initially are wired to randomly chosen neighbours, so that $G$ is an Erdos-Renyi directed graph at time $t=0$. As for the dynamics, let the state of each site be defined in terms of an integer variable $s_i\in\{0,1,2,\ldots,K_{out}^i\}$.
Given a configuration $\vec s$ at time $t$, the dynamics proceeds as follows:
\begin{itemize}
  \item Pick a site at random and add one unit (token) to its variable $s_i\to s_i+1$
  \item If $s_i<K_{out}^i$ for all sites $i$, then $t\to t +1$ and go back to the previous step
  \item If $s_i\ge K_{out}^i$ for any site $i$, distribute its units to its downstream neighbours:
  \begin{eqnarray*}
s_i & \to & 0\\
s_j & \to & s_j+g_{i,j}\qquad\forall j
\end{eqnarray*}
\end{itemize}
i.e., $K_{out}^i$ also acts as the threshold for toppling. After the first set of unstable ($s_i\ge K_{out}^i$) sites are relaxed following the above dynamics, we check for other sites which may have become unstable because of receiving units from its unstable neighbours, and relax them following the same procedure. We keep performing these toppling waves untill all sites of the network are stable. We call such an event between time $t$ and time $t+1$ an {\em avalanche}. On a random network we expect the statistics of avalanche sizes and duration to follow well known behaviours \cite{LubeckHeger,MunozZapperi}. Note that any site having more units than $K_{out}^i$ will dissipate the excess units, thereby maintaining a steady state for the total number of units (tokens) in the network.

\subsection{Network dynamics: Hebbian Dynamics}

\subsubsection{\bf{The Parametric Model}}

The dynamics we implement promotes co-activation of neighbouring sites. If $g_{i,j}=1$ and $j$ becomes active with large probability once it receives an input from $i$, then the link $i\to j$ is maintained. Instead, if $j$ stays idle with large probability, the link $i\to j$ will be rewired to a different site $j'$. In order to implement this, let $n_{i,j}$ be the number of inputs that $i$ sent to $j$ since the establishment of the link $i\to j$, and let $k_{i,j}$ be the number of times that site $j$ became active following a perturbation at site $i$ (clearly $k_{i,j}\le n_{i,j}$). A Bayesian estimate of the probability with which $j$ will fire next time it receives an input from $i$ is given by\footnote{The probability to have $k_{i,j}$ co-activation events, given $n_{i,j}$ inputs from node $i$ to $j$, can be expressed as a binomial distribution 
\[
P\{k_{i,j}|n_{i,j},\mu_{i,j}\}={ n_{i,j}\choose k_{i,j}} \mu_{i,j}^{k_{i,j}}(1-\mu_{i,j})^{n_{i,j}-k_{i,j}}
\]
where $\mu_{i,j}$ is the probability of a co-activation event on link $i\to j$. The distribution of $\mu_{i,j}$, given $n_{i,j}$ and $k_{i,j}$ can be obtained by Bayes rule, assuming a prior distribution $p_0(\mu_{i,j})$ for $\mu_{i,j}$. The expected value over this distribution yields Eq. (\ref{muij}), with the choice of an uniform prior $p_0(\mu_{i,j})=1$ for $\mu_{i,j}\in [0,1]$.}
\begin{equation}
\label{muij}
\langle \mu_{i,j}|n_{i,j},k_{i,j}\rangle =\frac{k_{i,j}+1}{n_{i,j}+2}.
\end{equation}
At the end of each avalanche, a number $L$ of links are chosen at random. For each of them if $\langle \mu_{i,j}|n_{i,j},k_{i,j}\rangle < \mu$ then the link $i\to j$ is ``rewired'' to a different site $i\to j'$, where $j'$ is chosen at random. Note that here the total number of links in the network is conserved. Also, the number of outgoing links is fixed for all the sites.
Depending on the value of $\mu$ (for $L$ finite) we expect different phases. For $\mu$ small we expect that most of the links will not satisfy the condition and hence will not be rewired. The model will not explore efficiently the space of network architectures, so the initial random network will prevail. For $\mu$ close to one, instead we expect that $L$ links are rewired very frequently, so also in this case, the network remains a random directed graph. For intermediate values of $\mu$ we may expect an intermediate phase where some links become more stable than others, as a consequence of the self-organisation of the flow of perturbations. This should be the interesting phase. 

\subsubsection{\bf{The Extremal Model}}

Next, we study a dynamics in which instead of rewiring $L$ links with $\langle \mu_{i,j}|n_{i,j},k_{i,j}\rangle < \mu$, we select, after each avalanche, the one link with the lowest value of $\langle \mu_{i,j}|n_{i,j},k_{i,j}\rangle$ and rewire it from $i\to j$ to a different site $i\to j'$, where $j'$ is again chosen randomly. In section \ref{sec:extremal}, we show that a random network with such extremal dynamics self-organises to the intermediate non-trivial phase of the parametric model.

\subsubsection{\bf{The Adaptive Model}}
\label{sec:modfatigue}
The dynamics of the steady state in the clustered network is such that the hubs are active almost at all timesteps. From a neurophysiological perspective this is not plausible, since the typical firing rate of a neuron is far below its upper bound, given by the inverse of the refractory period. In fact, adaptation mechanisms, such as \emph{neural fatigue} (based on a limited amount of available synaptic resources \cite{tsodyks1997,abbott1997}) prevent the neuron from having a very high firing rate.
Therefore, to avoid fixation in the network, we introduce
 a threshold number of topplings $N_f$ after which a site loses all its in-coming links. The rationale behind this choice is that the activity of a neuron with an excessively large firing rate would be homeostatically suppressed, thus lowering its coactivation with its upstream units, and ultimately, making its incoming links unstable. 
This mechanism could also be a more direct implementation of neuronal turnover which is seen in the olfactory bulb, where old neurons regularly die (along with all their connections) and are replaced with new neurons, 
or in social networks where it could mimic the death of a leader.

The incoming links of the excessively active hub are then rewired to other random sites, conserving the total number of links. The fatigued site can subsequently gain and maintain future in-coming links, like any other site. In this sense we are coarse graining the timescale separation between the fatigue and recovery process of a site. The rest of the dynamics on, and of the network remains the same as in Extremal Model.

Depending on the value of $N_f$ we expect different network architectures to appear, with small $N_f$ giving rise to a random, ``liquid" network and very large $N_f$ (of the order of the simulation length) resulting in the maintenance of the self-organised state of the extremal dynamics. Here we show that intermediate values of $N_f$ result in an interesting regime whose macroscopic properties are potentially relevant from a biological perspective.

\section{Results}

\subsection{Network undergoing parametric rewiring dynamics evolves towards a clustered state for a critical value of $\mu$}

\begin{figure}[h!]
\begin{center}
\includegraphics[width=0.75\linewidth]{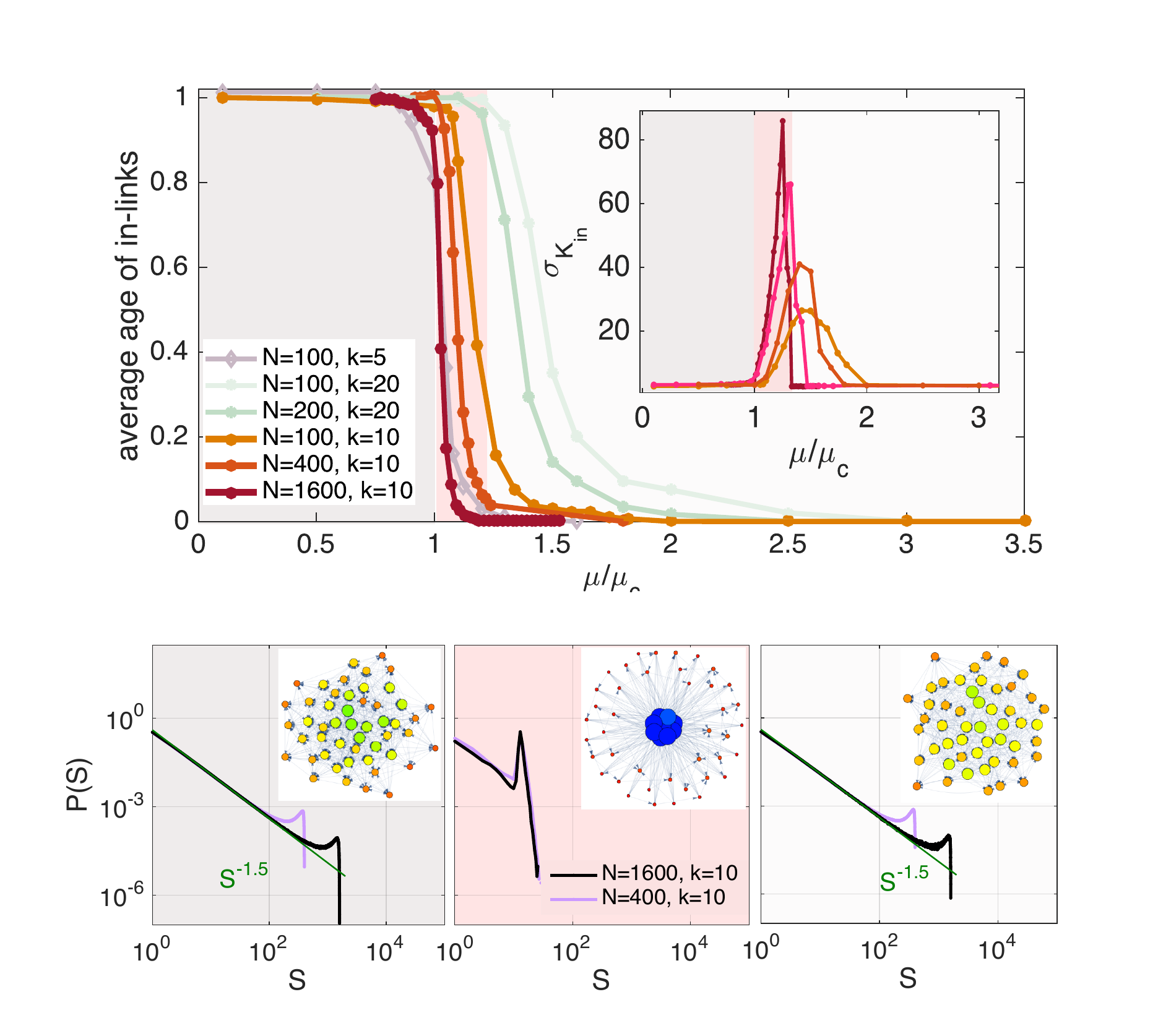}
\caption{\label{ptm} \textbf{Network undergoing parametric rewiring dynamics evolves towards a clustered state for a critical value of $\mu$:} Mean age of the in-coming links (top panel) and the standard deviation of the in-coming degree distribution (inset of top panel) as a function of $\mu$ shows clustering at $\mu = \mu_c$. The data points are obtained averaging across snapshots of the network taken every $9*10^5$ time steps (of the slow dynamics as defined in \ref{sec1}) in simulations of total length $10^9$ and excluding a transient of $10^8$ time steps. Bottom panels show avalanche size distributions and typical snapshots of a network in sub-critical (left panel, $\mu/\mu_c = 0.8$), critical (center panel, $\mu/\mu_c = 1.1$) and supercritical (right panel, $\mu/\mu_c = 1.5$) regimes.}
\end{center}
\end{figure}

The network dynamics for a fixed threshold $\mu$ of the rewiring rule (Parametric Model) are summarised in Fig. \ref{ptm}. In the top panel we report the average age, defined as the fraction of the simulation time over which a link  has remained unchanged. For small values of $\mu$ the links are very stable, whereas for large values of $\mu$ links are rewired frequently. As a result, in both extreme cases, the system is characterised by a random network (see Fig. \ref{ptm}, bottom panels), and its dynamics conforms with the typical  distribution of avalanche sizes\footnote{The size of an avalanche is measured as the number of sites that underwent at least one toppling event during the avalanche} for mean-field models~\cite{LubeckHeger,MunozZapperi}, which follow a power law behaviour with exponent $-1.5$. For an intermediate critical value of $\mu$, however, the network evolves towards a clustered state. In this state a few ``hub" sites receive incoming links from a finite fraction of all nodes, while others lose most of their incoming links. Consequently, the variance of the in-degree distribution peaks in this state (see inset in Fig. \ref{ptm}), and it diverges when $N\to\infty$. The avalanche size distribution exhibits a bump, corresponding approximately to the number of hubs, followed by a sharp cut off. This happens because the activity typically reverberates among the hubs,  in the clustered state. 

Note that all the sites have the same number of out-going links ($K_{out}^i = k$) and hence also the same threshold for toppling throughout the whole dynamics. This symmetry among sites is spontaneously broken with the increase in clustering and the formation of hubs. This symmetry breaking is promoted by the fluctuation in the number of incoming links. 

We find that the critical value of the threshold is given by $\mu_c=1/k$, which is approached in the limit of large system size with finite $k$. This can be understood by the following argument: The probability that the addition of a random unit causes a downstream site to topple is $1/k$, because the  threshold for toppling is $k$. Hence we expect that $k_{i,j}$ grows with $n_{i,j}$ approximately as $k_{i,j}\approx n_{i,j}/k$. Therefore, for $\mu< 1/k$ the number of rewired links will be very small whereas if $\mu> 1/k$ a large number of links will be rewired.

The formation of hubs, in the critical regime\footnote{Note that the criticality that is referred to here is pertinent to the dynamics \emph{of} the network, not to be confused with the SOC dynamics \emph{on} the network.}, can also be understood by the fact that once a node gathers an excess of incoming links, it will be more likely to topple during an avalanche, with respect to a node with few incoming links. At the same time, the activity concentrates over nodes with an excess of incoming links, which are those that are more likely to receive inputs. This twin effect 
confers stability to links impinging on sites with an excess of incoming links. This results in a positive feedback loop, by which nodes with an excess of incoming links will retain these links, and acquire new ones from the rewiring of unstable links impinging on nodes with a shortfall of incoming links. In this way, a weak inhomogeneity in the initial distribution of incoming links will be amplified by the dynamics and ultimately lead to the formation of hubs.

These results hold even when $L=N$, i.e., if all links with $\langle \mu_{i,j}|n_{i,j},k_{i,j}\rangle < \mu$ are rewired.

\subsection{Network undergoing extremal dynamics evolves towards the critical clustered state.}
\label{sec:extremal}

\begin{figure}[h!]
\begin{center}
\includegraphics[width=\linewidth]{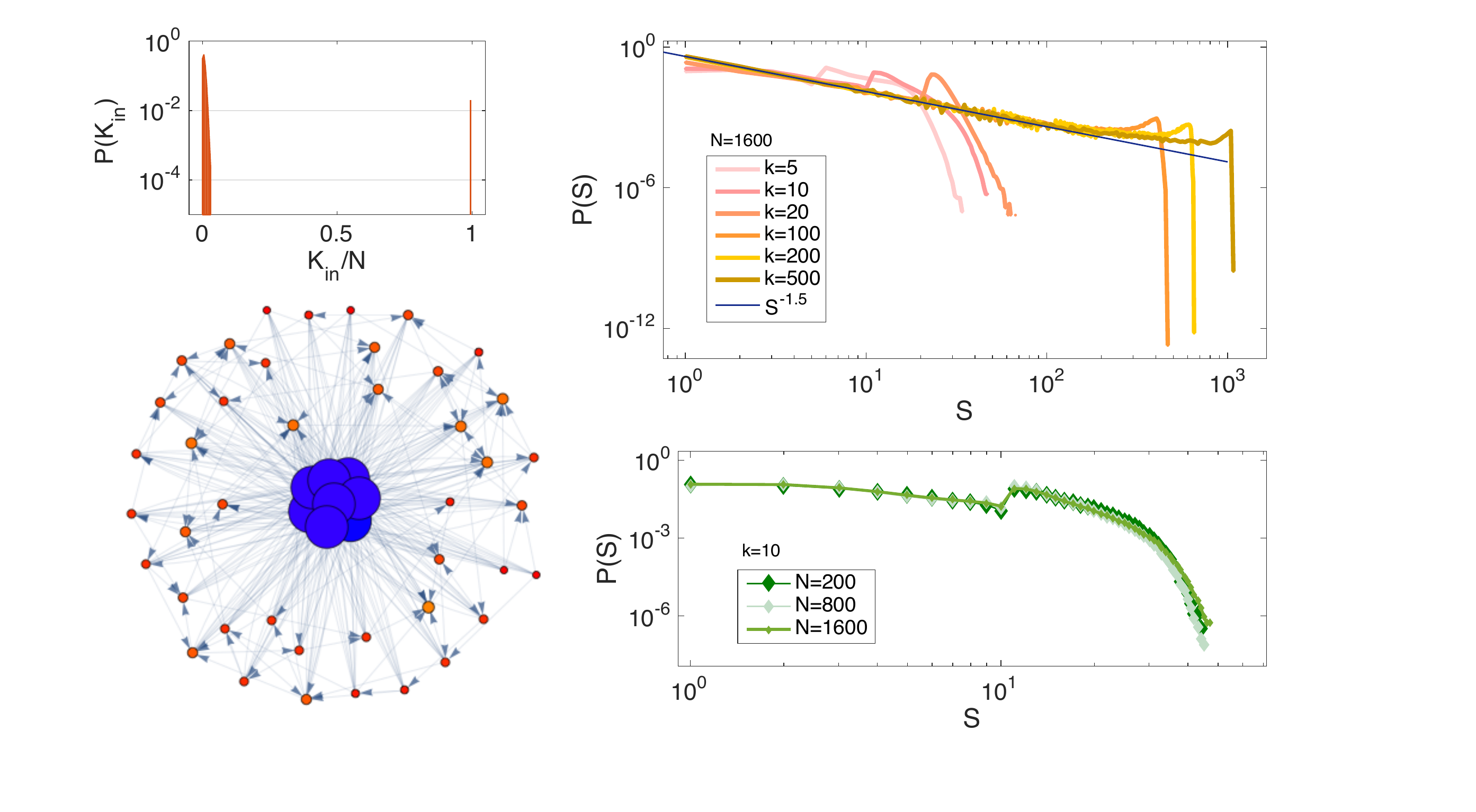}
\caption{\label{eso} \textbf{Network undergoing extremal dynamics evolves towards the critical clustered state:} The in-coming degree distribution (top-left panel) and a typical steady-state snapshot (bottom-left panel) demonstrates clustering under extremal dynamics (color code and the size of each node are a representation of their in-degree). The in-degree distribution is obtained averaging across snapshots of the network taken every $10^6$ time steps (of the slow dynamics as defined in \ref{sec1}) in simulations of total length $2*10^9$ and excluding a transient of $10^9$ time steps. The avalanche distribution (top right) exhibits a peak, corresponding to the number of hubs, followed by sharp cut-off.}
\end{center}
\end{figure}

The arguments given above for the formation of clustered network remain valid also when we adopt an extremal dynamics for the rewiring of links. This means that, instead of rewiring all links satisfying Eq. \ref{muij}, only the link with lowest value of $\langle \mu_{i,j}|n_{i,j},k_{i,j}\rangle$ is rewired after every avalanche.
As shown in Fig. \ref{eso}, extremal dynamics allows the system to reach the critical point with no need of parameter fine-tuning. Fig. \ref{eso} shows that the network under extremal dynamics evolves towards a clustered state with the same features as the critical state discussed above for the Parametric Model. We find that the number of hubs is very close to $k$, in most simulations, and the out-going links of hubs are directed to other hubs. Hence, toppling of one of the hubs, causes toppling events in other hubs, thereby confining the dynamics on the subpopulation of hubs. A signature of this is that the peak in the avalanche distribution is proportional to $k$ and it is independent of the system size (see Fig. \ref{eso}). 
We note that for $k\rightarrow N$ we recover the mean-field case in which the clique is as big as the network and the avalanche size exponent approaches the mean field value of $-1.5$. 

\subsection{Introduction of fatigue leads to dynamic disappearance and re-appearance of hubs}

\begin{figure}[h!]
\begin{center}
\includegraphics[width=\linewidth]{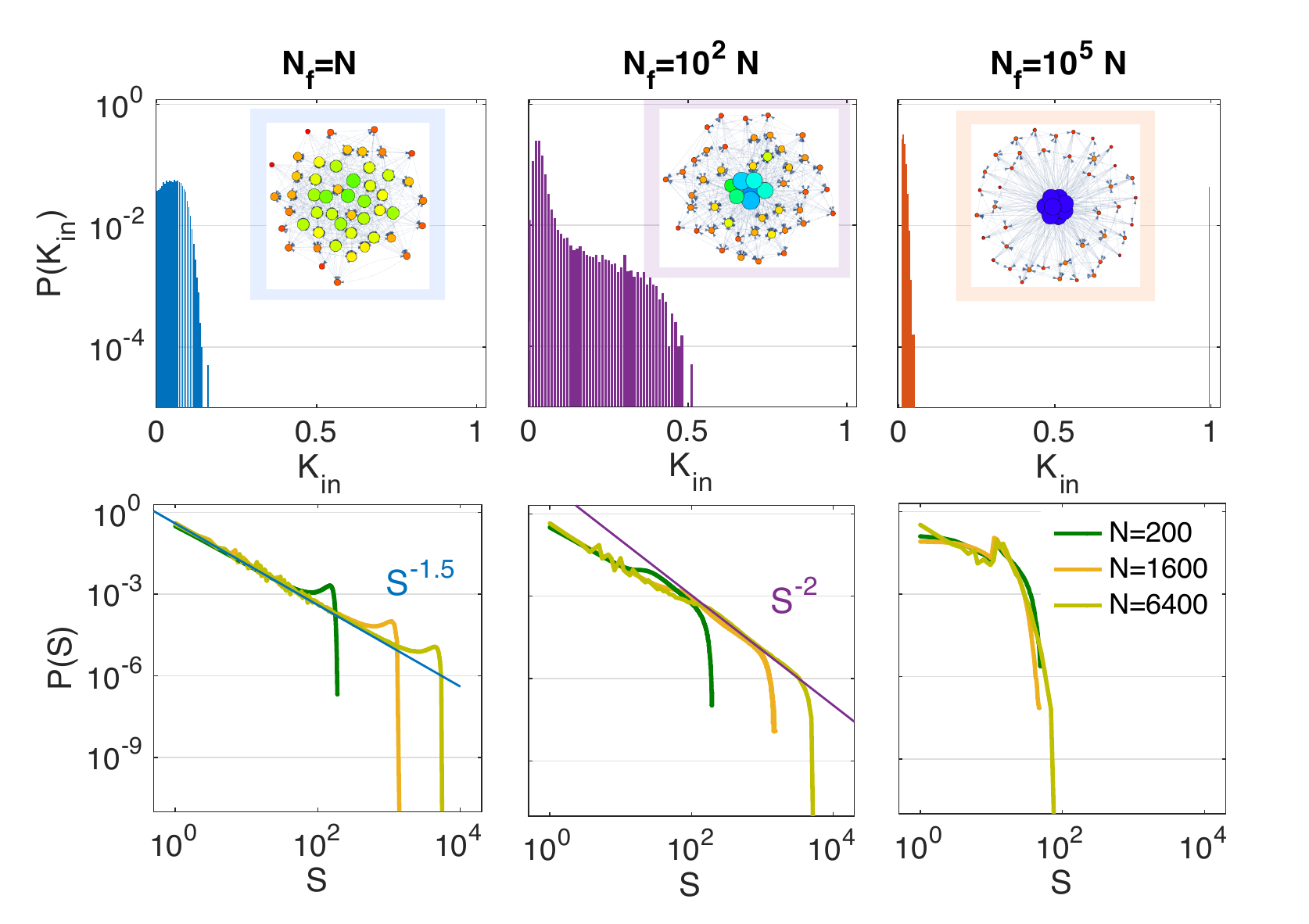}
\caption{\label{fat} \textbf{Network with fatigue leads to dynamic disappearance and re-appearance of hubs, depending upon the threshold for fatigue $N_f$:} The in-coming degree distributions (top panels), along with typical steady-state snapshots (top panel insets) and avalanche size distributions (bottom panels) show that the stability of the hubs depend upon the threshold for fatigue $N_f$, with small value leading to a random network (left panels) and large value leading to a clustered network (right panels). Interestingly, at intermediate value we see an intermediate network where avalanche distribution shows non-trivial power-law which is not SOC.}
\end{center}
\end{figure}

The clustered state reached in the evolution of Parametric Model, with $\mu=\mu_c$ or in the Extremal Model, is very robust against a variety of modifications in the details of the algorithm.
Nevertheless, as discussed in Section \ref{sec:modfatigue} a clustered network may not be most versatile for information processing. The very existence of a sharp cutoff in the avalanche size distribution limits information spreading to $s\sim\mathcal{O}(k)$ sites.
 Furthermore, as argued above, realistically the system would implement some homeostatic mechanism to prevent such an inefficient distribution of processing power. Therefore, we shall now introduce a mechanism mimicking the effect of fatigue,
in order to see how it modifies the network and the avalanche dynamics.
Fatigue is mediated by the parameter $N_f$, which is the number of firing events after which the incoming links of a site are reset: as expected, a small value of $N_f$ prevents the formation of any hub and the network remains random (Fig. \ref{fat}). In fact, hubs, which are the sites with the highest number of toppling events, are the nodes most affected by the fatigue mechanism. For very large values of $N_f$ the effect of fatigue are negligible and the system's behavior stays unchanged with respect to the self-organized clustered state of the extremal dynamics. Interestingly, for intermediate values of $N_f$, the system evolves towards a non-trivial network configuration, with non vanishing probability of finding avalanches as big as the network size and a power law avalanche size distribution extending over few decades with an exponent compatible with $-2$ (as e.g. in \cite{marsili1998self}, see also \cite{yaghoubi2018} and \cite{pozzi2020}).

\section{Discussion}

This paper investigates the type of structures that emerge from simple rules that couples the activity of a network of nodes, to the evolution of the network itself. We explore local self-reinforcement rules reminiscent of the synaptic plasticity of neurons \cite{Hebb,turrigiano2004homeostatic}, whereby the synapses between two neurons is reinforced when the post-synaptic neuron is activated by a pre-synaptic signal. Our results suggest a natural tendency of neural networks to form strongly connected cores of active nodes surrounded by passive ones. The activity in the resulting network concentrates strongly on the core nodes (the hubs). These results are rather robust and persist even when we limit the activity of nodes by taking into account the effects of fatigue on very active nodes. 
These results are in agreement with previous work showing that unconnected neurons whose activity is correlated tend to develop a new connection, while uncorrelated neighbors tend to disconnect \cite{bornholdt2003}, but go a step further, showing that such mechanism is self-reinforcing, hence leading to the formation of a dense core. 
Moreover, it has been shown that in a dynamical network of excitatory and inhibitory units in which coupling strength co-evolves with the dynamics on the node, optimal choice of adaptive dynamics leads to clustering \cite{singh2014}.

The network architectures produced by the stylised models presented here are very far from the striking complexity and diversity of architectures found in real neuronal circuits, already in simple organisms such as {\em C. elegans}~\cite{varshney2011structural}. Still, the structure we find is reminiscent of the insect brain, where many electrically passive neurons are arranged around the periphery, with the neural signal is carried by an active core of neurons \cite{turner2016insect} 

Also, our results are reminiscent of the ``long-lasting transient periods of increased firing at individual sites"  observed in neural networks grown in vitro  \cite{van2005dynamics}. 

Similar self-reinforcement rules can also be at play in shaping social networks. The cost of a social link may not be compensated by a sufficient reward, when the recipient of the link is idle or non-responsive. It is tempting to speculate that a mechanism similar to the one described here may be responsible for the peculiar statistical properties observed in social networks  \cite{leskovec2008statistical}.

\section*{References}
\bibliographystyle{iopart-num}
\bibliography{biblio}

\end{document}